# Structural modulation and BC8 enrichment of silicon via dynamic decompression

Yubing Du,[1,2] Guoshuai Du,[1,2] Hongliang Dong,[3] Jiayin Li,[1,4] Wuxiao Han,[1,2] Bin Chen,[3] and Yabin Chen[1,2,5,*]

[1]*Advanced Research Institute of Multidisciplinary Sciences, Beijing Institute of Technology (ARIMS), Beijing 100081, P.R. China*

[2]*School of Aerospace Engineering, Beijing Institute of Technology, Beijing 100081, P.R. China*

[3]*Center for High Pressure Science and Technology Advanced Research, Shanghai 201203, P.R. China*

[4]*School of Chemistry and Chemical Engineering, Beijing Institute of Technology, Beijing, 100081 P. R. China*

[5]*BIT Chongqing Institute of Microelectronics and Microsystems, Chongqing, 400030, P. R. China*

[*]Correspondence and requests for materials should be addressed to: chyb0422@bit.edu.cn (Y.C.)

**ABSTRACT:** The modern very large-scale integration systems based on silicon semiconductor are facing the unprecedented challenges especially when transistor feature size lowers further, due to the excruciating tunneling effect and thermal management. Besides the common diamond cubic silicon, numerous exotic silicon allotropes with outstanding properties can emerge under high pressure, such as the metastable BC8 and metallic *β*-tin structures. Despite much effort on the controlled synthesis in experiment and theory, the effective approach to rationally prepare Si phases with desired purity is still lacking and their transition mechanism remains controversial. Herein, we systematically investigated on the complicated structural transformations of Si under extreme conditions, and efficiently enriched BC8-Si phase via dynamic decompression strategy. The splendid purity of BC8-Si was achieved up to ~95%, evidently confirmed by Raman spectroscopy and synchrotron X-ray diffraction. We believe these results can shed a light on the controlled preparation of Si metastable phases and their potential applications in nanoelectronics.

**INTRODUCTION**

Silicon (Si) as a key material in semiconductors and nanoelectronics, has drawn the extensive attention for several decades [1, 2]. Diamond cubic Si (Si-I, space group $Fd\bar{3}m$) along with its fundamental properties have been well investigated under ambient condition [3-7]. Extreme conditions, such as hydrostatic pressure, can arouse numerous phase transitions, discover the extraordinary Si structures, and further stimulate many practical applications [8-10]. High-pressure researches on Si started more than 60 years ago, and since then plenty of novel Si allotropes have been reported by using diamond anvil cell (DAC) [6, 11, 12].

Phase transitions of Si under extreme conditions are inconceivably complicated. During compression process, Si can undergo a series of phase transitions from diamond cubic Si-I at ambient condition to the tetragonal $\beta$-Sn phase (Si-II, space group $I4_1/amd$) at ~11.7 GPa [13, 14], to the orthorhombic symmetry (Si-XI, space group $Imma$) at ~13.2 GPa, and to a simple hexagonal structure (Si-V, space group $P6/mmm$) at ~15.4 GPa [15]. Notably, all high-pressure phases of Si-II, Si-XI, and Si-V have been proved to be metallic in experiments, and thus their electrical conductivity and carrier density exceed those of Si-I in principle [16, 17]. In comparison, during decompression process, transition pathway of Si-II →Si-XI →Si-V is reversible, that is, each phase can be recovered under its critical pressure. Importantly, when pressure lowers further, instead of a reversible transition to the original Si-I, Si-II phase transforms to a rhombohedral R8 phase (Si-XII, space group $R\bar{3}$) at ~9.4 GPa [18], and sequentially to a metastable body-centered BC8 phase (Si-III, space group $Ia\bar{3}$) [19]. It is reported that both Si-XII and Si-III are kinetically stable, and the latter can be reserved till ambient pressure. Moreover, transmission measurements demonstrated that Si-III is narrow-gap semiconductor with a tiny band gap $E_g$ ~30 meV [20], while Si-XII is an indirect semiconductor and $E_g$ approximates 0.24 eV [21]. More interestingly, hexagonal diamond (HD) phase (Si-IV, space group $P6_3/mmc$) emerges once Si-III is thermally annealed at a moderate temperature ~470 K [22], and the indirect band gap 0.95 eV of Si-IV is comparable to that of Si-I [23]. When temperature goes up to 1050 K, the thermodynamically stable Si-I phase comes back, accompanied by the fully relaxed lattice structure [24]. Alternatively, Si-IV phase can transform to Si-II at ~11.2 GPa. Fig. 1(a) summarizes the detailed transition pathways of Si under various temperatures and pressures.

Among those exotic metastable phases in Fig. 1(b), Si-III emerges these years and has drawn much attention due to its potential applications in optoelectronics and photovoltaics [25, 26]. To prepare the pure Si-III phase is with long-term challenge, owing to its co-existence with Si-XII once pressure is released, even though it has been discovered since 1963 [27]. In theory, optical band gap of Si-III remains controversial, severely relying on the used theoretical methods. For instance, Cohen, et al. found that the indirect band gap of Si-III is calculated as ~0.43 eV [21], while the obtained band gap via transmittance measurement reached as narrow as ~30 meV [20]. The present methods to prepare Si-III impose the significant limitations on sample size and co-existed Si-III/Si-XII as mixture [28]. Therefore, there is still lack of efficient method to prepare the pure Si-III phase. It is reported that the remaining fraction of R8-Si was dramatically reduced by thermal annealing [29]. Recently, the pure polycrystalline BC8-Si was tentatively synthesized under 14 GPa and 900 K [30]. Moreover, despite much effort on theoretical simulation, the transition mechanics between metastable Si phases remains ambiguous. In terms of transition dynamics, the synergetic effect of multi factors principally governs the ultimate Si structure and properties, including compression rate, temperature, and pressure and shearing stress [31].

In this work, we have attempted to rationally modulate the phase structures of Si, and further to disclose the transition mechanism through designing the various kinetic processes. It is found that metastable Si-III phase can be significantly enriched up to ~95% via the dynamic decompression strategy. The slow decompression rate can efficiently suppress the formation of R8-Si, evidenced by both Raman spectra and synchrotron X-ray diffraction results. In addition, the metallic Si-II phase can be retained as low as ~ 5 GPa, when the decompression process occurs at 80 K. These results can provoke many intriguing pathways to regulate the metastable structures of Si-based semiconductors and other functional materials.

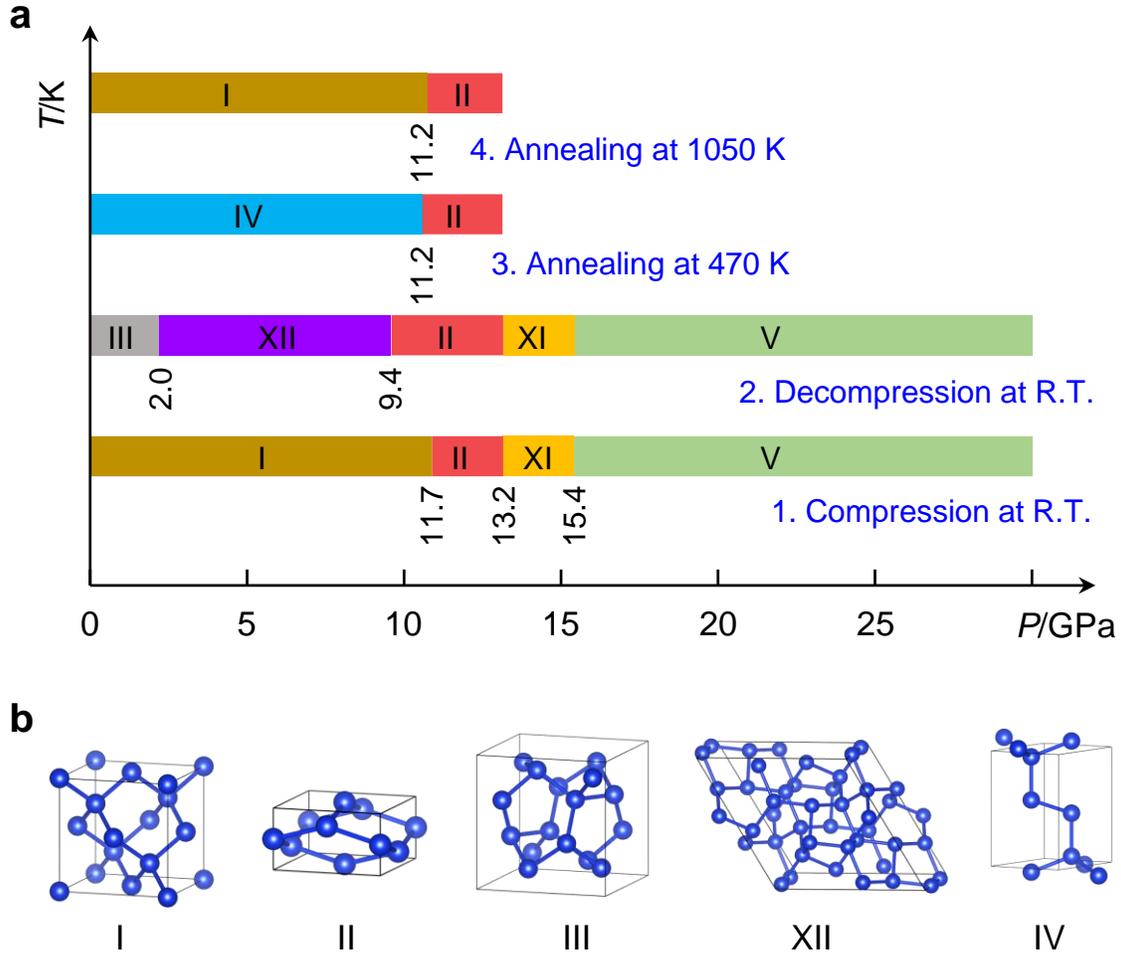

**Figure 1.** Schematic diagram of phase transitions and lattice structures of Si. (a) Phase transitions of silicon under high temperature and high pressure, including the compression and decompression processes. Notably, Si-I to Si-II transition is irreversible. Si-IV can be obtained via thermal annealing at ~470 K, and it transforms back to Si-I at higher temperature like ~1050 K. Each number labeled in the transition pathways indicates the critical pressure. (b) The lattice structures of Si allotropes (from left to right): diamond cubic Si-I, tetragonal Si-II, body-centered Si-III, rhombohedral Si-XII, and hexagonal Si-IV.

## RESULTS AND DISCUSSION

To gain more insights about the pressure effect on lattice dynamics of Si, *in situ* Raman measurements of the ultrathin Si crystal were performed under various pressures as shown in Fig.

2. During the compression process, the representative Raman spectra in Fig. 2(a) shows that the first-order $F_{2g}$ mode of Si-I phase vibrates at 520.7 cm$^{-1}$ at 0 GPa, originated from the degenerate transverse optical (TO) and longitudinal optical (LO) phonon at Γ point in Brillouin zone, and it remarkably shifts to larger wavenumber as pressure. The extracted slope d$\omega$/d$P$ via linear fitting is 4.3 cm$^{-1}$/GPa, consistent with the literature results [32]. In contrast, the second-order Raman mode of Si-I from two transverse acoustic (2TA) phonons presents a significant red-shift with pressure, from ~303 cm$^{-1}$ at 0 GPa to 250.9 cm$^{-1}$ at 9.2 GPa, primarily owing to the pressure-induced softened behavior. At ~11.4 GPa, both $F_{2g}$ and 2TA vibrations disappear with the undetectable intensity, suggesting Si-I to Si-II transition. It is obvious that two weak Raman peaks at 121.4 and 385.9 cm$^{-1}$ (under 11.4 GPa) rose up, attributed from the LO and TO phonons of metallic Si-II based on group theory, respectively. More interestingly, these LO and TO modes exhibit the distinct pressure-dependence in Fig. 2(b): the hardened LO phonon vs softened TO phonon under high pressure, possibly due to the anisotropic tetragonal lattice as reported previously [33].

In comparison, during decompression process, the LO and TO phonons of Si-II phase reversibly vibrate first and then become inapparent till ~9 GPa [19, 34], accompanied with the emergent Si-XII phase as discussed above. The typical Raman peaks of Si-XII lie at 165.1, 351.7, and 397 cm$^{-1}$ in Fig. 2(a), corresponding to two $A_g$ modes and one $T_u$ mode. Intriguingly, Raman intensity of these modes is significantly enhanced with the released pressure, implying the sluggish transition process from Si-II to Si-XII. When pressure is below ~2.5 GPa, Si-III phase gradually appears, partly transformed from Si-XII, and its proportion approach to the extremum under ambient condition. It is emphasized that Raman spectrum of Si-XII concomitantly exist with that of Si-III, due to their comparable atomic structures, as shown in the lower panel of Fig. 2(a). In terms of lattice symmetry, the rhombohedral Si-XII structure can be considered as a distorted Si-III phase, resulting in eight Si atoms in its own unit cell [35]. Table 1 summarizes the Raman shifts and calculated Grüneisen parameters ($\gamma$) of various Si phases, including the six vibrational modes of Si-III and other six modes for Si-XII. More importantly, the Si-XII to Si-III transition was evidently proved by an apparent slope change of Raman shift versus pressure, i.e., from 1.8 to 1.2 cm$^{-1}$/GPa at ~2 GPa for 350 cm$^{-1}$, as illustrated in Fig. S1-S2 and 2(c).

**Table 1.** Lattice structures, phonon modes, Raman shifts, and Grüneisen parameters of the various Si allotropes.

| Structure | Phonon mode | Raman shift (cm$^{-1}$) (this work) | Raman shift (cm$^{-1}$) (references) | $\gamma$ (this work) | $\gamma$[36, 37] |
|---|---|---|---|---|---|
| Si-I (fcc) | 2TA(X) | 303.7 | 301.9[38, 39] | -1.5 | |
| | TO | 520.7 | 520.3[38, 39] | 0.44 | 0.42 |
| Si-II (Tetragonal) | LO | 121.4 | ~121[33] | | |
| | TO | 385.9 | ~386[33] | | |
| Si-III (BC8) | T$_g$ | | 182.4[40] | | |
| | - | | 373.3[41] | | |
| | T$_g$ | 385.3 | 384.2[40, 42] | 0.22 | 0.24 |
| | T$_u$ | 407 | 412[40, 42] | 0.14 | |
| | E$_u$ | 434.3 | 437.5[40, 42] | 0.76 | 0.81 |
| | E$_g$ | | 463[19] | 1.14 | 1.42 |
| Si-XII (R8) | A$_g$ | 165.1 | 164.8[19] | -0.44 | -0.26 |
| | E$_g$ | | 170.0[19] | | |
| | A$_g$ | 351.7 | 351.9[19, 36, 42] | 0.65 | 0.46 |
| | T$_u$ | 397 | 397.1[19, 36, 42] | 0.45 | 0.28 |
| | A$_g$ | | 440.3[19] | | |
| | A$_u$ | | 453.6[19] | | |
| a-Si (Amorphous) | TA | ~147 | ~150 (broad)[43] | | |
| | | ~289 | 300 | | |
| | | ~373 | 390 | | |
| | TO/LO/LA | ~468 | ~470 (broad)[43] | | |

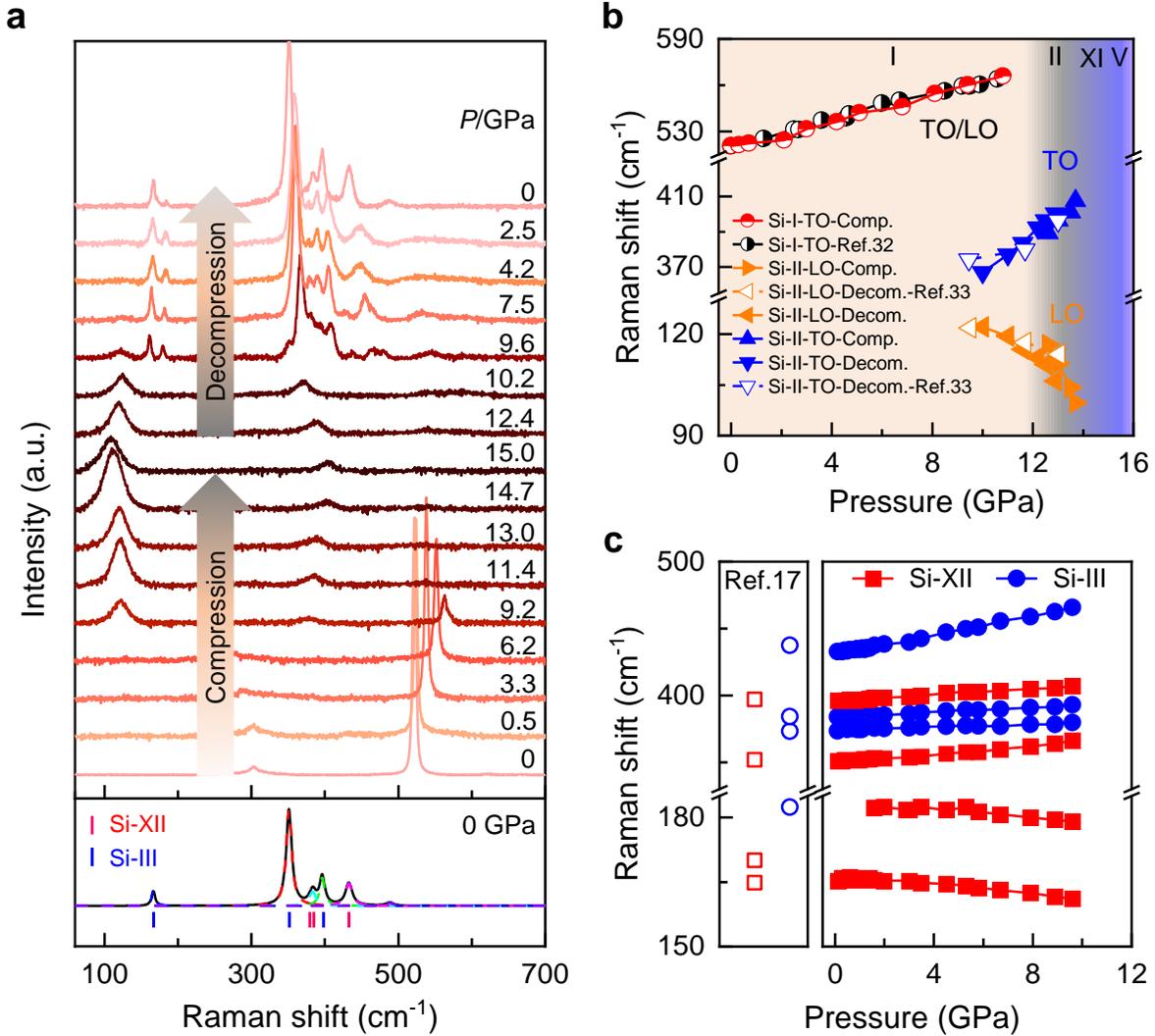

**Figure 2.** Representative Raman spectra and the irreversible phase transitions of Si under high pressure. (a) Pressure-dependence of Raman spectra for Si allotropes up to 15 GPa. The Si-I transforms to metallic Si-II around 11.4 GPa, and to Si-XII and Si-III consequently once the pressure released (upper part). The fitting result of Raman spectrum of the totally decompressed Si-III and Si-XII (lower part). The blue and red tick marks denote the Raman modes of Si-III and Si-XII, respectively. (b) The extracted Raman shifts of Si-I (note that LO and TO phonons are degenerated) and Si-II as a function of pressure. The Raman peaks remain almost unchanged after compression and decompression processes, and the literature results are concluded for comparison as well. (c) Raman shift evolution of Si-XII and Si-III during decompression process, together with the calculated results from Ref. 17.

To obtain high-purity Si-III, the dynamic decompression strategy has been demonstrated by considering the diverse Si phases as precursors, with respect to their transition kinetics [44]. The rapid unloading rate is realized via combining gas membrane and piezoceramics compression technique [12] while the slow decompression is operated manually, spanning five orders of magnitude from $\sim 5.0\times10^{-4}$ to $\sim 50$ GPa/s as shown in Fig. 3(a). Manifestly, the Raman mode $\sim 434.3$ cm$^{-1}$ (at 0 GPa) of Si-III and $\sim 397$ cm$^{-1}$ (at 0 GPa) of Si-XII exhibit the distinct and remarkable changes, that is, the former one becomes stronger significantly while the latter inversely weaker as unloading rate increases, leading to the enhanced ratio of $I_{Si-III}/I_{Si-XII}$. The $I_{Si-III}/I_{Si-XII}$ data as a function of decompression rate displays a clear exponential relationship, and can be well fitted quantitatively with $y = 0.38x^{-0.06}$ in Fig. 3(b). Following this trend, it can be assumed that the pure Si-III phase can be reasonably achieved with ultraslow decompression rate, which unfortunately exceeds the capability of common setups. In Fig. 3(c), the desired purity of Si-III was determined to approach $\sim 95\%$ using our optimized decompression recipe, as proved by the synchrotron X-ray diffraction (XRD) characterization. Regarding to the complicated transition mechanism, it can be preliminarily understood that the slow decompression rate allows more thorough nucleation centers of Si-III phase, and the transformation pathway is synchronously mediated with their activation barrier and free energy in terms of thermodynamics.

Morevoer, the various precursors obtained at specific high pressure are alternatively considered to modulate Si-III to Si-XII ratio. As shown in Fig. 3(d) and 3(e), we investiagted four high-pressure Si allotropes as precursors, inculding Si-II at 12 GPa, Si-XI at 14 GPa, Si-V at 22 GPa, and Si-IV at 12 GPa. The initial Si-IV phase was acquired by thermally annealing Si-III at $\sim 573$ K, totally confirmed by their Raman features. As discussed above, it is feasible to quantitatively analysis the Si-III proportion in the coexisted Si-III/Si-XII by Raman spectrum [45]. Evidently, each curve of Raman shift versus pressure exhibits an apprant slope change. For example, the lienar slope (430 cm$^{-1}$) changes from $\sim 4.5$ to 4.1 cm$^{-1}$/GPa around $\sim 2$ GPa for Si-IV preccuror in Fig. 3d. Regardless of the original phase strucutre, it corresponds to Si-XII to Si-III transition. The detailed Raman spectra of various Si allotropes are depicted in Fig. S3 and S4. Importabtly, it is obvious that Si-II preceursor is beneficial to Si-III formation, and the obtained $I_{Si-III}/I_{Si-XII}$ reached ~0.6. This phenomenon can be relatively explained by the reversible transition between Si-II and Si-III/XII, compared with other precursor phases, which thus eventually favors the Si-III once decompression.

Next, we turn to the temperature effect on the transition behavior from Si-II to Si-XII and Si-III during decompression process, as shown in Fig. 4. When decompression happened at 80 K, Si-II can exist till 5.7 GPa, dramatically lower than ~10.1 GPa at which Si-II was decompressed at room temperature. With the pressure released further, it is obvious that two board and week Raman bands at ~147 and ~468 cm$^{-1}$ appear, suggesting the formation of a-Si state [46, 47]. This transition can be explained by the fact that the reduced thermal energy at 80 K can't overcome activation barrier from Si-II to Si-XII/III, and thus the a-Si with lower free energy was retained. The pressure-dependent Raman shifts of Si-II in Fig. 4(b) confirm that the slopes of both TO (10.8 cm$^{-1}$/GPa) and LO (-5.2 cm$^{-1}$/GPa) modes at 80 K are comparable to that at room temperature. This result implies that the pure Si-II could be potentially acquired when decompressed at even lower temperature. Furthermore, we also investigated the influence of high temperature on phase transitions of Si during unloading process. The high-temperature and high-pressure condition realized by laser-heated DAC acted to Si-II phase at ~12 GPa, and the temperature calibrated through black-body radiation was 1200 K (near to the melting point of Si), or 1900 K at which Si became liquid. It is evident that the $I_{Si-III}/I_{Si-XII}$ ratio seems relatively larger when heated at 1900 K, as presented in Fig. 4(c). More detailed Raman results at 1200 K and 1900 K are shown in Fig. S5. This exotic behavior was observed in Si nanowires as well [48], which is related to rearrangements of Si atoms from liquid to crystalline phases.

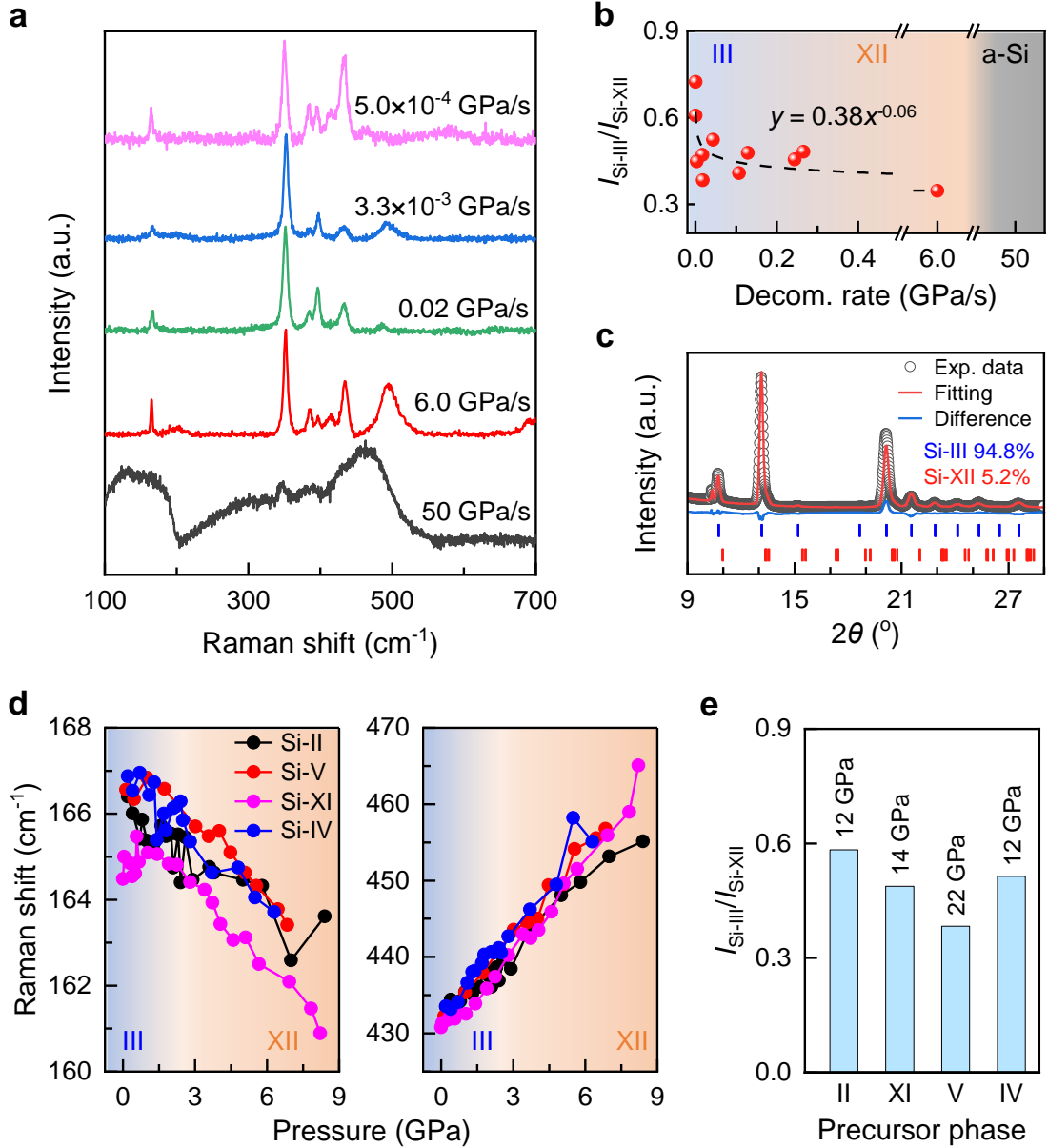

**Figure 3.** The significant enrichment of Si-III via the dynamic decompression process. (a) Raman spectra of Si samples obtained under different decompression rates, varying from ~5.0×10$^{-4}$ to ~50 GPa/s. (b) $I_{Si-III}/I_{Si-XII}$ varied as decompression rate. $I_{Si-III}$ and $I_{Si-XII}$ imply Raman intensity of 430 cm$^{-1}$ for Si-III and 395 cm$^{-1}$ for Si-XII. The red data can be fitted exponentially, as denoted by dashed line. Gray area means amorphous Si (a-Si). (c) Synchrotron XRD curve of Si-III/Si-XII mixture acquired by the optimized decompression rate, and purity of Si-III reached ~94.8%. (d) Raman frequency evolution of metastable Si-III and Si-XII decompressed from different precursors. (e) The obtained $I_{Si-III}/I_{Si-XII}$ when decompressed from Si-II, Si-XI, Si-V, and Si-IV.

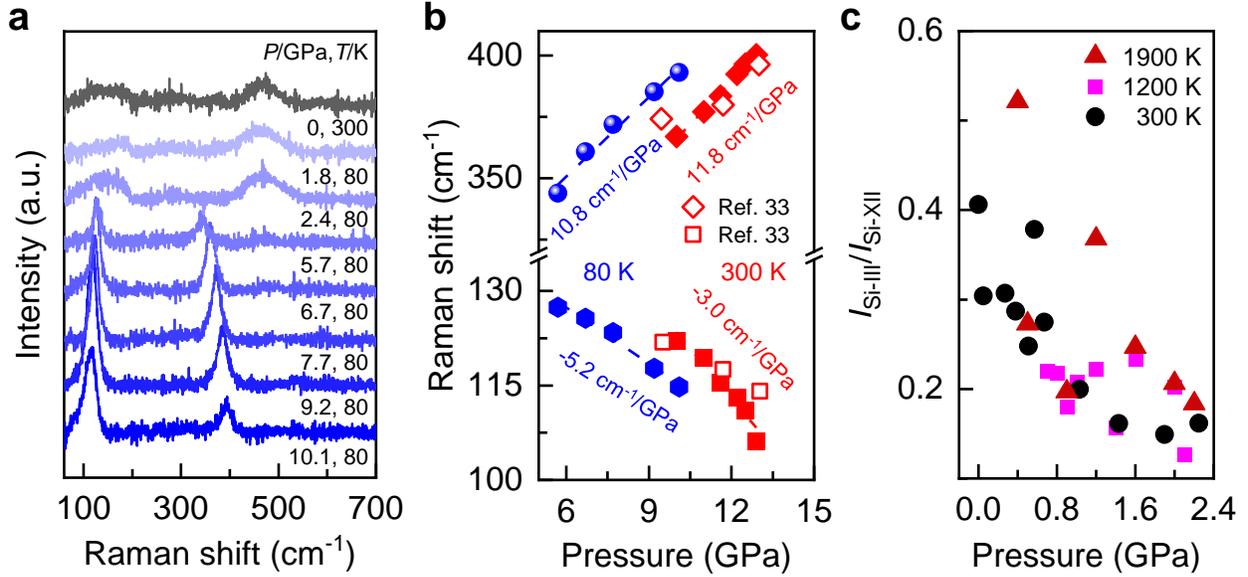

**Figure 4.** Temperature effect on the structural modulation of Si phases. (a) Pressure-dependent Raman spectra of Si when decompressed at 80 K (blue). The gray curve was measured at 300 K, and the final product is characterized as a-Si. (b) Raman shift evolution of Si-II during unloading process at 80 K (blue) and 300 K (red). The slopes via linear fitting (dashed lines) are labeled as well. (c) Pressure-dependence of Raman intensity ratio of Si-III (430 cm$^{-1}$) to Si-XII (395 cm$^{-1}$). The Si samples were quenched from 1900 (dark red), 1200 (pink), and 300 (black) K.

To explore the distinguished properties of Si under extreme conditions, we performed the pressure-dependent photoluminescence (PL) characterizations of Si to regulate its band gap and phonon energies. As shown in Fig. 5(a), the acquired PL spectra of Si at 80 K are featured with a broad band and several week shoulder peaks, owing to the indirect band gap characteristic of Si-I phase. The band gap energy $E_g$, TO and TA phonon energies at 80 K, extracted through Gaussian fitting of experimental data, are determined as 1.16 eV, 60.9, and 35.8 meV, respectively, well consistent with literature results [49-51]. Compared with the obtained results at 300 K in Fig. S7, the obvious blue-shift is originated from the enhanced electron-phonon couplings as temperature decreases. The intensity of each PL band became weak, and eventually disappeared at ~ 4.5 GPa. As displayed in Fig. 5(b), pressure-dependent phonon energy shows a positive slope, *i.e.*, 1.6 and 2.3 meV/GPa for TA and TO modes at 80 K, respectively, which are relatively lower than 3.8 and 3.2 meV/GPa at 300 K. Furthermore, the band gap energy of Si is dramatically reduced with pressure as illustrated in Fig. 5(c), and the extract linear slops correspond to -15.8 and -8.5 meV/GPa at 80 and 300 K, respectively. Their negative correlation is ascribed from both the lifted X point of conduction band and descended Γ point of valence band in the band structure of Si-I phase. Apparently, the overall band gap energy is apparently enlarged at lower temperature, resulted from the suppressed phonon population.

**CONCLUSION**

In summary, an effective approach via dynamic decompression process was established to rationally modulate the phase structures of Si. We found that the ultraslow decompression rate can benefit to enrich the metastable BC8-Si from the co-existed BC8 and R8 mixture. X-ray diffraction and Raman spectrum proved that the optimized purity of the obtained BC8-Si approached up to ~95%. Moreover, the metallic Si-II phase can be retained as low as ~5 GPa (two times smaller than that at 300 K) when the decompression process occurred at 80 K. Our work can pave the way for the phase-pure synthesis of Si-based semiconductors, and offer the novel clues to promote their applications in nanodevices and optoelectronics.

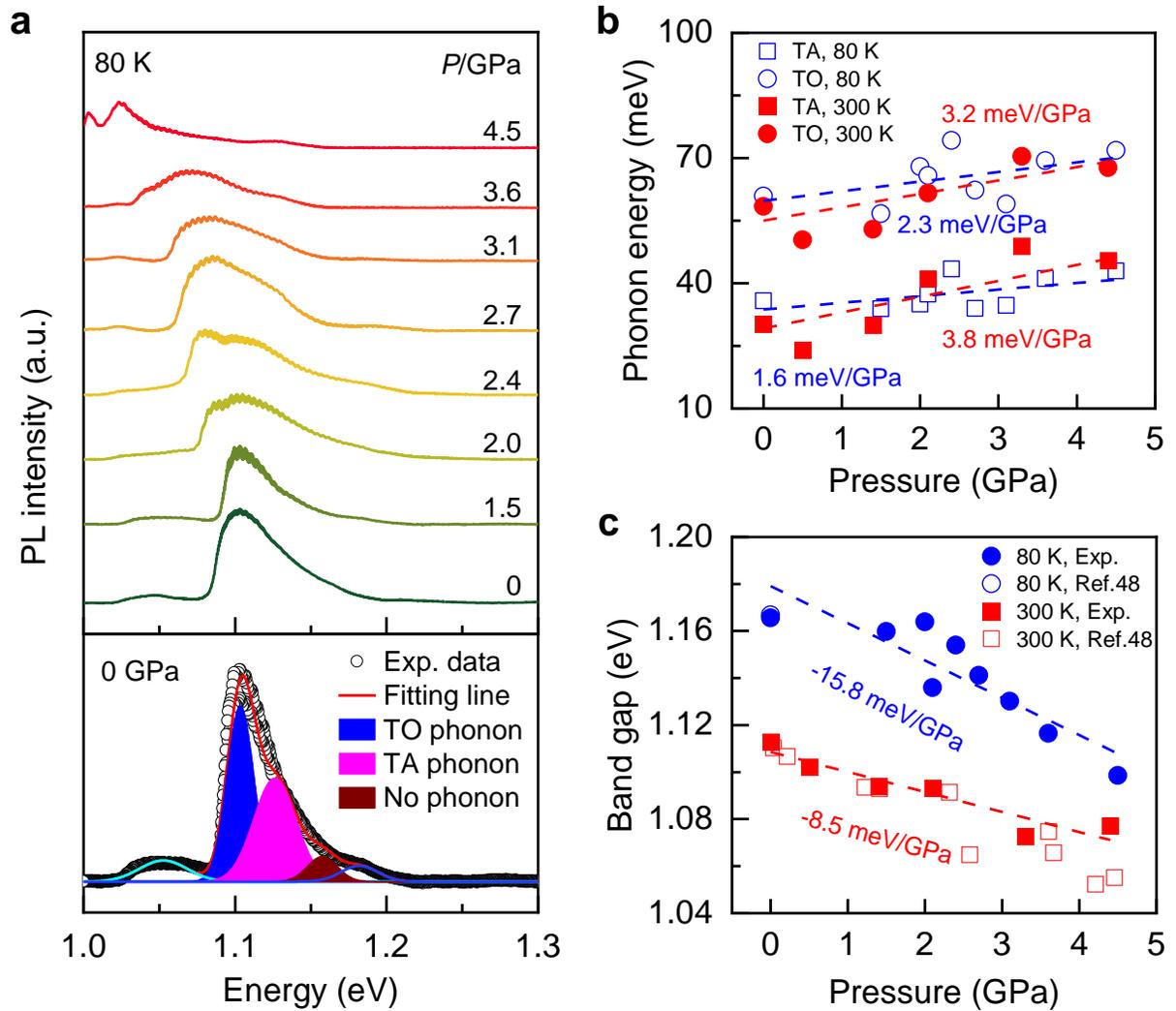

**Figure 5.** Pressure modulation of the exotic optical properties of Si at 80 K. (a) The obtained PL spectra of Si-I under the variable pressures at 80 K (upper part). The PL signal became undetectable at ~4.5 GPa. The fitting of PL spectrum at 0 GPa was performed with Gaussian function (lower part). (b) Energy evolution of both TA and TO phonons of Si-I under different pressures at 80 and 300 K. (c) Band gap energy evolution of Si under different pressures. The blue and red symbols denote the experimental data points at 80 and 300 K, respectively. The dashed lines represent the linear fitting results.

## EXPERIMENTAL METHODS

**Sample preparation for high pressure characterizations.** The metastable Si with various phases was prepared using a symmetric DAC with ~400 μm anvil culet. The sample chamber (~ 150 μm diameter hole) was drilled at the center of T301 stainless steel gasket with the pre-indented thickness of ~50 μm. The crystalline Si-I chip (~40 μm thick) was loaded into gasket chamber together with a tiny ruby ball. Pressure was calibrated by monitoring the significant shift of R1 fluorescence line of ruby under pressure. Mixture of methanol-ethanol-water (16: 3: 1) was served as pressure transmitting medium. For laser heating experiments, the gasket hole was filled partially with the pre-dried NaCl powder as pressure transmitting medium and the embedded silicon chip in the middle, in order to exclude the well-known pyrolysis effect of the common organic mediums. In this way, the Si sample was held apart from both diamond surface and gasket wall. When the applied pressure was above ~12 GPa, the metallic Si samples were heated up to desired temperature via the home-made laser heating system. The temperature was calibrated by fitting the acquired black-body radiation curve.

***In situ* Raman spectrum under high pressure.** *In situ* Raman spectra of Si phases under hydrostatic pressure were collected through the transparent diamond anvil of DAC by using Horiba iHR550 spectrometer, and the spectral resolution was superior to 0.1 cm$^{-1}$, which is offered by 1800 gmm$^{-1}$ grating. The wavelength of excitation laser was 632.8 nm, and the diameter of focused beam approached ~1 μm with 50× objective. The DAC can be conveniently loaded into the home-made cryostat setup for *in situ* Raman measurements under variable temperatures and pressures, and the minimum temperature approximated 77 K via liquid nitrogen. The pressure of DAC can be easily tuned under a specific low temperature.

**The synchrotron XRD measurements.** The crystal structures of the synthesized Si samples were characterized by the advanced synchrotron XRD at the beamline 15U1 of Shanghai Synchrotron Radiation Facility (SSRF). The diameter of the focused X-ray beam is around 3 μm and its wavelength is 0.6199 Å. Two-dimensional diffraction images were collected using a MAR165

CCD detector, and then integrated to one dimensional XRD curves through Dioptas software [52]. The XRD data was further analyzed and refined based on Rietveld method with GSAS-II [53].

## ASSOCIATED CONTENT

**Supporting Information**

The Supporting Information is available free of charge online.

Detailed Raman spectra of Si with cyclic loading-unloading processes; Raman spectra obtained by using different Si; pressure-dependent Raman spectra of Si samples after laser heating treatments; thermal stability of Si-IV; energy gap and phonon energy evolution of Si samples under compression at 300 K.

**Author Contributions**

Y.C., and Y.D. conceived this research project and designed the experiments. Y.D. prepared the samples and carried out Raman and PL measurements under high pressure. G.D., and W.H. attributed to Raman characterizations and *in situ* low-temperature experiments. H.D., and Y.D. performed the synchrotron XRD measurements, with supervision from B.C. J.L., and Y.D. performed the laser-heating experiments. Y.C., and Y.D. wrote the manuscript with the essential input of other authors. All authors have given approval of the final manuscript.

**Notes**

The authors declare no competing financial interests.

**Acknowledgements**

This work was financially supported by the National Natural Science Foundation of China (grant numbers 52072032, and 12090031), and the 173 JCJQ program (grant No. 2021-JCJQ-JJ-0159).

**Data Availability**

All data related to this study are available from the corresponding author on reasonable request.

**GRAPHICAL TABLE OF CONTENTS**

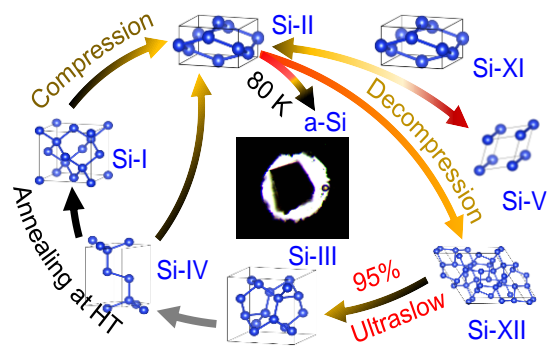

Pressure-induced structural modulations of silicon were systematically investigated under extreme conditions. Importantly, dynamic decompression strategy was rationally developed to significantly enrich BC8-Si, which can be extended to discover more novel allotropes.